\documentclass[aps,pra,amssymb, amsfonts,amsmath,showpacs, nofootinbib,superscriptaddress]{revtex4-1}
 \usepackage{graphicx}
  \usepackage{amssymb}
  \begin{document}
\title{Dynamical collapse for photons}
\author{Philip Pearle}
\email{ppearle@hamilton.edu}
\affiliation{Emeritus, Department of Physics, Hamilton College, Clinton, NY  13323}
\begin{abstract}
 {I suggest a simple alteration of my CSL (Continuous Spontaneous Localization) theory, replacing the mass density collapse-generating operators
by relativistic energy density operators.  Some consequences of the density matrix evolution equation are explored.  First,  the expression for the mean energy increase of free particles is calculated (which,  
in the non-relativistic limit, agrees with the usual result). Then, the density matrix evolution is applied to photons. 
The mean rate of loss of photon number from a laser beam pulse, 
the momentum distribution of  the photons ``excited" out of the laser beam pulse, and the alteration  of the cosmic blackbody spectrum are all treated to first order in the collapse rate parameter $\lambda$. Associated possible experimental limits on $\lambda$ are discussed.}

 \end{abstract}

\maketitle

\section{Introduction}\label{} 

Some time ago, I proposed the idea of a stochastic dynamical collapse theory\cite{p76+}, where a term which  depends upon a randomly fluctuating quantity is added to Schr\"odinger's equation.  As a result, a superposition of states (in a particularly chosen basis) is continuously driven toward one such state, with (neglecting the usual Hamiltonian evolution) the Born probability. 

In the CSL (Continuous Spontaneous Localization) theory\cite{pearle, gpr, reviews}, the randomly fluctuating quantity is a classical scalar field, and the term added to
 the Schr\"odinger equation depends as well upon a 
``collapse-generating operator." Initially, I chose this to be the particle number density operator\cite{pearle}, but later\cite{ewen} replaced it by the mass density operator so that the collapse is toward a mass density eigenstate. 

In addition to this modified Schr\"odinger equation, CSL is completed with the specification of the ``probability rule," that the probability of a given fluctuating field is proportional to the squared norm of the state vector which evolved under that field.

An important aspect of CSL collapse behavior  is that the collapse is very slow for micro-objects, but fast for macro-objects, behavior which was first embodied in the thereby justly celebrated Spontaneous Localization (SL) theory of Ghirardi, Rimini and Weber\cite{grw} (where, however, the evolution is discontinuous: also, fermion or boson wave function symmetry  is destroyed in SL, but a version which removes that flaw exists\cite{sdt}).  As a result, particle behavior is scarcely affected but, since we see macro-objects, ``what you see is what you get" from the theory.  

Events are common physical occurrences.   Standard quantum theory predicts the probabilities of 
 events but does not describe their occurrence: like Moses, it indicates the promised land, but does not go there. 
Standard quantum theory may therefore be justifiably regarded as incomplete: CSL may be regarded as providing a completion.  

Since photons do not have mass, in the present non-relativistic CSL theory (which has been, and is currently, the object of experimental scrutiny),  
photons do not contribute to collapse dynamics.  It does not seem that there is a physical reason why this should be so.  Since photons are relativistic particles, perhaps that has been waiting on the construction of a convincing, viable relativistic version of CSL\cite{rel}.  Until that happens, I propose the following. In non-relativistic CSL, replace the mass density operators $\xi^{\dagger}({\bf x})\xi({\bf x})$ with the energy density operators $[K^{1/2}\xi^{\dagger}({\bf x})][K^{1/2}\xi({\bf x})]$, where  
$K^{2}\equiv-\nabla^{2}+M^{2}$, and $\xi({\bf x})$ is the annihilation operator of a particle of mass $M$ at location ${\bf x}$.  Setting $M=0$ then gives 
collapse dynamics for (one polarized species of) photons. 

Another obvious possible choice of energy density operators is $\frac{1}{2}\xi^{\dagger}[K\xi({\bf x})] +\frac{1}{2}[K\xi^{\dagger}({\bf x})]\xi({\bf x})$.  
Both operators are Hermitian and their integral over all space gives the free particle relativistic Hamiltonian, the two basic requirements for the energy density operator.  
I have chosen to work with the one-term expression rather than the two-term expression simply because its square (which appears in the density matrix evolution equation) 
is one term, while the square of the other is more cumbersome, four terms.  Whether there is a physical reason for preferring one over the other I do not know, 
nor have I looked to see how the other choice might affect the calculations in this paper. 

Of course, this is not a relativistically invariant theory, although collapse caused by differences in relativistic energy density does capture aspects  
of what one could expect in such a theory. One might consider the proposal here as representing collapse in the preferred, co-moving frame.  

An important difference between the usual mass density operators and the energy density operators  is that the 
commutator of the former operators at any two spatial points vanishes, while this is not so for the latter operators.
In the former case, this allows one to make use 
of a theorem that,  if all collapse-generating operators mutually commute,  there is collapse toward  
the mutual joint eigenstates of these operators (neglecting the Hamiltonian evolution).  Thus, one is assured of collapse toward mass density eigenstates  in the former case.

For the latter case, one does not have that easy assurance.  However, 
at least for massive particles, the commutator is quite small\footnote{
$[K^{1/2}\xi^{\dagger}({\bf x})K^{1/2}\xi({\bf x}),K'^{1/2}\xi^{\dagger}({\bf x}')K'^{1/2}\xi({\bf x}')]=
\Big(K^{1/2}\xi^{\dagger}({\bf x})K'^{1/2}\xi({\bf x}' )-K'^{1/2}\xi^{\dagger}({\bf x}')K^{1/2}\xi({\bf x})\Big)K^{1/2}K'^{1/2}\delta({\bf x}-{\bf x}')$
is the commutator, and  $K^{1/2}K'^{1/2}\delta({\bf x}-{\bf x}')=\frac{1}{(2\pi)^{3}}\int d{\bf k}\sqrt{k^{2}+M^{2}}e^{i{\bf k}\cdot({\bf x}-{\bf x}')}
=\frac{1}{2\pi^{2}|{\bf x}-{\bf x}'|}\int_{0}^{\infty}kdk\sqrt{k^{2}+M^{2}}\sin k|{\bf x}-{\bf x}'|=-\frac{M^{2}}{2\pi^{2}|{\bf x}-{\bf x}'|^{2}}
\Big[K_{0}(M|{\bf x}-{\bf x}'|) +\frac{1}{M|{\bf x}-{\bf x}'|}K_{1}(M|{\bf x}-{\bf x}'|)\Big]$.  The result above follows from the large argument approximation of the 
Bessel function $K_{0}$.},
 although not vanishing, $\rightarrow-[(2\pi)^{3}\lambdabar_{M}^{3}|{\bf x}-{\bf x}'|^{5}]^{-1/2}e^{-|{\bf x}-{\bf x}'|/\lambdabar_{M}}$, for $|{\bf x}-{\bf x}'|/\lambdabar_{M}>>1$  
($\lambdabar_{M}\equiv \hbar/Mc$ is the reduced Compton wavelength of the particle).  This suggests looking for an extension of the theorem to ``almost" commuting 
operators, which shall not be pursued here.   

Instead, one may  look at examples, to see how collapse dynamics evolves. The basic requirement  of a collapse theory is that, 
when one considers a superposed state  of many particles in two different places, there is collapse toward all particles being in one or the other place.
An example is given in Appendix \ref{A}, where the particles are moving, so that  relativistic behavior may come into play.  There, for the state $|\psi,0\rangle=\frac{1}{\sqrt{2}}[|L\rangle+|R\rangle]$, the two spatially displaced states $|L\rangle, |R\rangle$ each consist of N  particles,
each particle in the same state occupying a volume $\sim \sigma^{3}$,  
each particle moving with well-defined momentum ${\bf k}_{0}$ in a direction orthogonal to their displacement vector ${\bf x}_{L}-{\bf x}_{R}$.  In this example, the resulting density matrix behavior describing energy density generated collapse turns out to be identical to that when there is mass density generated collapse, except that 
the collapse rate factor $\sim M^{2}$ is replaced by $\omega^{2}(k_{0})\equiv k_{0}^{2}+M^{2}$.  

We shall not review here the CSL dynamical equation for the state vector and how one derives the Lindblad equation for the density matrix from it and the probability rule\cite{ reviews,misc}.  We shall just start with that density matrix evolution equation, with the above substitution: 
\begin{equation}\label{1}
\frac{\partial}{\partial t}\rho(t)=-i[H,\rho(t)]-\frac{\lambda}{2M_{N}^{2}}\int d{\bf x}\int d{\bf x}'e^{-({\bf x}-{\bf x}')^{2}/4a^{2}}
[K^{1/2}\xi^{\dagger}({\bf x})K^{1/2}\xi({\bf x}),[K^{1/2}\xi^{\dagger}({\bf x}')K^{1/2}\xi({\bf x}'), \rho(t)]],
\end{equation}
and proceed from there. (Here, $M_{N}$ is the mass of the neutron, $\lambda$ is the collapse rate and $a$ the collapse range, typically chosen as the SL suggested values $\lambda\approx 10^{-16}$s and $a\approx10^{-5}$cm, but limits on these phenomenological constants are being experimentally pursued.)

Applying (\ref{1}) to the collapse example mentioned above,  the off-diagonal density matrix element between the 
two states, to order $\lambda$, is given by Eq.(\ref{A10}):
\begin{eqnarray}\label{A10}
\langle L|\rho(t)|R\rangle 
&\approx&\frac{1}{2}-\frac{N\lambda t\omega^{2}(k_{0})}{2M_{N}^{2}}\Big[N\Big(\frac{a}{\sigma}\Big)^{3}\Big(1-e^{-({\bf x}_{L}-{\bf x}_{R})^{2}/4\sigma^{2}}\Big)+1\Big]\nonumber
\end{eqnarray}
\noindent This clearly describes the decay of the matrix element, for any values of the parameters consistent with the assumptions underlying (\ref{A10}), 
$k_{0}\sigma>>k_{0}a>>1$.  

In this paper, we shall discuss  the explicit collapse behavior no further than this example calculated in  Appendix A.  For we are particularly interested in the ``anomalous" excitation of photons ($M=0$) 
which is a byproduct of the collapse dynamics.  (By ``anomalous" is always meant behavior not predicted by standard quantum theory, and therefore open to experimental test.)
 
Because collapse narrows wave functions, the momentum and therefore the energy of particles is ``anomalously" increased.  In the non-relativistic theory based upon mass density-generated collapse, the rate of energy increase of $N$ identical non-relativistic particles  is\cite{ewen}
\begin{equation}\label{2}
\frac{d}{dt}\bar H=\lambda\frac{3\hbar^{2}}{4Ma^{2}}\frac{M^{2}}{M_{N}^{2}}N.
\end{equation}
\noindent In Section II, the comparable relativistic expression shall be obtained from Eq.(\ref{1}) (with (\ref{2}) as the non-relativistic limit). 

In Sections III, IV, we consider the effect of collapse on a beam or pulse of laser light. The state vector is a coherent state, a superposition of states of various numbers of photons of almost identical momentum, where the number of photons obeys Poisson statistics. These states have different  energy densities. Insofar as the collapse dynamics tries to evolve the state vector toward one of these states, while this changes the  statistics for a single beam, 
it doesn't affect the Poisson statistics for the ensemble of beams because the collapse dynamics respects the Born rule. 

But, these states are expected to be modified since the collapse mechanism also imparts energy to photons,  which removes them from 
a coherent beam. That \textit{will} affect the statistics of the ensemble, and decrease the mean number of photons in the beam.  In section III, we calculate the loss in the 
ensemble-mean number of photons from the laser beam, to first order in the collapse rate parameter $\lambda$. We apply the result to an experimentally achieved 
 intense laser beam pulse, which is in the infra-red, and also to an experimentally achieved  x-ray laser beam pulse, to see what upper limits on $\lambda$ could be implied.  

Photon number is conserved.   The photons lost from the beam are made more energetic by the collapse process. In Section IV we calculate the 
momentum distribution of these ``anomalous" photons. We consider how these photons are ejected from an experimentally achieved intense CW laser beam, again suggesting a limit on $\lambda$.  

Section V is motivated by the consideration that the longer the collapse process acts, the more photons are excited.  Therefore we discuss the cosmic blackbody photons, as they are affected by collapse 
over the time interval since recombination sent them freely on their way, almost over the age of the universe.  
There is an ensuing distortion of the blackbody spectrum, but  the resulting effect is small.  
This is partly because there are so few photons involved, $\approx 16\pi(kT/hc)^{3}\approx 400$photons/cc, and partly because their energy 
$\approx 2.5\times 10^{-6}$eV to $2.5\times 10^{-2}$eV  is so small (photons with wavelength 50cm to .05cm). 

 \section{Energy Increase}\label{II}
The mean energy of a collection of identical particles of mass $M$ described by the density matrix $\rho(t)$ is $\bar H(t)\equiv Tr H\rho(t)$, where 
$H\equiv\int d{\bf x}\xi^{\dagger}({\bf x})K\xi({\bf x})$, and $Tr$ is the trace operation.  Then by Eq.(\ref{1}),
\begin{eqnarray}\label{3}
\frac{\partial}{\partial t}\bar H(t)&=&-\frac{\lambda}{2M_{N}^{2}}Tr\rho(t)\int d{\bf x}\int d{\bf x}'\int d{\bf x}''e^{-({\bf x}-{\bf x}')^{2}/4a^{2}}\nonumber\\
&&[K^{1/2}\xi^{\dagger}({\bf x})K^{1/2}\xi({\bf x}),[K^{1/2}\xi^{\dagger}({\bf x}')K^{1/2}\xi({\bf x}'), \xi^{\dagger}({\bf x}'')K\xi({\bf x}'')]].  
\end{eqnarray}
\noindent Writing $\xi({\bf x})\equiv \frac{1}{(2\pi)^{3/2}}\int d{\bf k}a({\bf k})e^{i{\bf k}\cdot {\bf x}}$ (i.e., $a({\bf k})$ is the annihilation operator of a particle of momentum ${\bf k}$), Eq. (\ref{3}) 
becomes, with $\omega(k)\equiv\sqrt{k^{2}+M^{2}}$:

\begin{eqnarray}\label{4}
\frac{\partial}{\partial t}\bar H(t)&=&-\frac{\lambda}{2M_{N}^{2}}\frac{1}{(2\pi)^{6}}Tr\rho(t)\int d{\bf x}\int d{\bf x}'\int d{\bf k}_{1}d{\bf k}_{2}d{\bf k}_{3}d{\bf k}_{4}d{\bf k}_{5}
\sqrt{\omega(k_{1})\omega(k_{2})\omega(k_{3})\omega(k_{4})}\omega(k_{5})                                    \nonumber\\
&&e^{-({\bf x}-{\bf x}')^{2}/4a^{2}}e^{-i ({\bf k}_{1}-{\bf k}_{2})\cdot{\bf x}} e^{-i ({\bf k}_{3}-{\bf k}_{4})\cdot{\bf x}'}
[a^{\dagger}({\bf k}_{1})a({\bf k}_{2}),[a^{\dagger}({\bf k}_{3})a({\bf k}_{4}),a^{\dagger}({\bf k}_{5})a({\bf k}_{5})]]  \nonumber\\
&=&-\frac{\lambda}{2M_{N}^{2}}\frac{1}{(2\pi)^{3}}(4\pi a^{2})^{3/2}Tr\rho(t)\int d{\bf k}_{1}d{\bf k}_{2}d{\bf k}_{3}d{\bf k}_{4}d{\bf k}_{5}
\sqrt{\omega(k_{1})\omega(k_{2})\omega(k_{3})\omega(k_{4})}\omega(k_{5})                                    \nonumber\\
&&e^{-({\bf k}_{1}-{\bf k}_{2})^{2}a^{2}}\delta(-{\bf k}_{1}+{\bf k}_{2}-{\bf k}_{3}+{\bf k}_{4})\nonumber\\
&&[a^{\dagger}({\bf k}_{1})a({\bf k}_{5})\delta({\bf k}_{4}-{\bf k}_{5})\delta({\bf k}_{2}-{\bf k}_{3})
-a^{\dagger}({\bf k}_{3})a({\bf k}_{2})\delta({\bf k}_{4}-{\bf k}_{5})\delta({\bf k}_{1}-{\bf k}_{5}) \nonumber\\
&&-a^{\dagger}({\bf k}_{1})a({\bf k}_{4})\delta({\bf k}_{3}-{\bf k}_{5})\delta({\bf k}_{2}-{\bf k}_{5})
+a^{\dagger}({\bf k}_{3})a({\bf k}_{2})\delta({\bf k}_{3}-{\bf k}_{5})\delta({\bf k}_{1}-{\bf k}_{4})]  \nonumber\\
&=&\frac{\lambda}{2M_{N}^{2}}\frac{1}{(2\pi)^{3}}(4\pi a^{2})^{3/2}2Tr\rho(t)\int d{\bf k}_{1}d{\bf k}_{2}a^{\dagger}({\bf k}_{1})a({\bf k}_{1})
\omega_{1}\omega_{2}(\omega_{2}-\omega_{1})
e^{-({\bf k}_{1}-{\bf k}_{2})^{2}a^{2}}
\end{eqnarray}
\noindent where, in the second step, the commutation operations have been performed and, in the last step, delta function integrals have been performed, and labels 1 and 2 have been exchanged in a term. 

In the non-relativistic limit, $\omega_{1}\omega_{2}(\omega_{2}-\omega_{1})\approx M^{2}[k_{2}^{2}-k_{1}^{2}]/2M$. The integral over ${\bf k}_{2}$ in (\ref{4}) is then 
\begin{equation}\label{5}
\int d{\bf k}_{2}[k_{2}^{2}-k_{1}^{2}]e^{-({\bf k}_{2}-{\bf k}_{1})^{2}a^{2}}=
\int d{\bf k}_{2}[({\bf k}_{2}-{\bf k}_{1})^{2}+2({\bf k}_{2}-{\bf k}_{1})\cdot{\bf k}_{1}]e^{-({\bf k}_{2}-{\bf k}_{1})^{2}a^{2}}=\frac{\pi^{3/2}}{a^{3}}\frac{3}{2a^{2}}.
\end{equation}
\noindent Inserting (\ref{5}) into Eq.(\ref{4}), we obtain Eq.(\ref{2}) ($N=Tr\rho(t)\int d{\bf k}a^{\dagger}({\bf k})a({\bf k}))$.

In the general case, we write  (\ref{4}) as
\begin{eqnarray}\label{6}
\frac{\partial}{\partial t}\bar H(t)&=&\frac{\lambda}{M_{N}^{2}}\Big(\frac{a^{2}}{\pi}\Big)^{3/2}Tr\rho(t)\int d{\bf k}_{1} a^{\dagger}({\bf k}_{1})a({\bf k}_{1})
f({\bf k}_{1}) \hbox{ where }\nonumber\\
f({\bf k}_{1})&\equiv& \int d{\bf k}_{2} \omega_{1}\omega_{2}(\omega_{2}-\omega_{1})
e^{-({\bf k}_{2}-{\bf k}_{1})^{2}a^{2}}
\end{eqnarray}

So far, Eq.(\ref{6}) is exact. 
We shall obtain analytic expressions in two approximate cases.

One is for photons when the density matrix $\rho$ is such that $ka<<1$ (here $k\equiv k_{1}$). Note that, with $a=100$nm, $2\pi a$ is in the neighborhood of red light's wavelength, so we are considering wavelengths  $\gtrsim$infrared. 

The other case is for  $ka>>1$. For photons, we are therefore considering wavelengths  $\lesssim$ultraviolet.  For electrons,  $ka>>1$ implies 
that the energy $(\hbar k)^{2}/2m_{e}>>(\hbar/a)^{2}/2m_{e}\approx 3\times 10^{-6}$eV, which 
of course means validity in a broad non-relativistic realm as well. 

For photons and $ka<<1$, in the integral  $f({\bf k}_{1})$, we write $\omega_{2}-\omega_{1}\approx k_{2}$ and  $({\bf k}_{2}-{\bf k}_{1})^{2}\approx k_{2}^{2}$, 
obtaining $f({\bf k}_{1})=k_{1}(\pi/a^{2})^{3/2}(3/2a^{2})$.

For $ka>>1$ we  change the variable of integration to ${\bf \Delta}\equiv {\bf k}_{2}-{\bf k}_{1}$. We expand $\omega_{2}$ to order $\Delta^{2}\sim1/a^{2}$, 
thereby omitting terms of order $(k_{1}a)^{-2}$ compared to the terms that are retained:  
\begin{eqnarray}\label{7}
f({\bf k}_{1})&=&\omega_{1}\int d{\bf \Delta}\sqrt{M^{2}+({\bf k}_{1}+{\bf \Delta})^{2}}[\sqrt{M^{2}+({\bf k}_{1}+{\bf \Delta})^{2}}   -\omega_{1}]e^{-{\bf \Delta}^{2}a^{2}}\nonumber\\
&\approx&\omega_{1}\int d{\bf \Delta}\Bigg[\omega_{1}+\frac{{\bf k}_{1}\cdot{\bf \Delta}}{\omega_{1}} \Bigg] 
\Bigg[ \frac{{\bf k}_{1}\cdot{\bf \Delta}}{\omega_{1}}+ \frac{{\bf \Delta}^{2}}{2\omega_{1}} - \frac{(2{\bf k}_{1}\cdot{\bf \Delta})^{2}}{8\omega_{1}^{3}}          \Bigg]e^{-{\bf \Delta}^{2}a^{2}} 
\nonumber\\
&=&\frac{\omega_{1}}{2}\int d{\bf \Delta}\Bigg[  {\bf \Delta}^{2}+ \frac{({\bf k}_{1}\cdot{\bf \Delta})^{2}}{\omega_{1}^{2}}\Bigg]
=\omega_{1}\frac{\pi^{3/2}}{a^{5}}\Bigg[\frac{3}{4}+\frac{k_{1}^{2}}{4\omega_{1}^{2}} \Bigg],  
\end{eqnarray}
\noindent so that 
\begin{eqnarray}\label{8}
\frac{\partial}{\partial t}\bar H(t)\approx\lambda\Big(\frac{\lambdabar_{N}}{a}\Big)^{2}Tr\rho(t)\int d{\bf k}_{1} a^{\dagger}({\bf k}_{1})a({\bf k}_{1})
\omega_{1}\Bigg[\frac{3}{4}+\frac{k_{1}^{2}}{4\omega_{1}^{2}} \Bigg]
\end{eqnarray}
\noindent where $\lambdabar\equiv\hbar/M_{N}c\approx 2\times10^{-14}$cm is the reduced Compton wavelength of the nucleon. 

Once again, we note that the non-relativistic limit Eq.(\ref{2}) is obtained from (\ref{8}), with $\omega_{1}\approx Mc^{2}, k_{1}/\omega_{1}\approx 0$.

Our two approximate expressions are therefore, first, for photons with  a density matrix $\rho$ describing photons such that $ka<<1$ and, second, from (\ref{8}), applicable both to massive particles in the relativistic regime 
 ($k_{1}/\omega_{1}\approx 1$)  and to photons ($k_{1}/\omega_{1}= 1$),  with $ka>>1$:  
\begin{subequations}
\begin{eqnarray}\label{9}
\frac{\partial}{\partial t}\bar H(t)&=&\lambda\frac{3}{2}\Big(\frac{\lambdabar_{N}}{a}\Big)^{2}\bar H(t) \hbox{ \qquad        for  $ka<<1$.}\label{9a}\\
\frac{\partial}{\partial t}\bar H(t)&=&\lambda\Big(\frac{\lambdabar_{N}}{a}\Big)^{2}\bar H(t) \hbox{ \qquad        for  $ka>>1$.}\label{9b}
\end{eqnarray}
\end{subequations}

The new wrinkle here is that there is exponential growth of the mean energy, not the non-relativistic linear growth  (\ref{2}). However, since over the age of the universe $T$, with $\lambda T\approx 40$, 
the exponent $\lambda T(\frac{\lambdabar_{N}}{a})^{2}\approx 10^{-16}$, the exponential growth is effectively linear and there is a negligible fractional contribution of collapse-induced 
energy to the universe.  

In spite of the smallness of this exponent, one should hasten to add that collapse-induced energy effects can have consequences that are not out of the realm of observability, since they can produce anomalous behavior, such as rare but unusual events, which may be experimentally singled out. Non-relativistically, 
this includes knocking electrons out of atoms\cite{atoms}, breaking up the deuterium nucleus\cite{sno}, shaking free charged particles so they radiate\cite{Fu}, inducing random walk in small objects\cite{ps}, contributing to the cosmological constant\cite{sud}. 

It is also worth emphasizing that, for massive particles, the relativistic result (\ref{8})  only applies to free particles.  However, the non-relativistic result (\ref{2}) is the energy increase even when there is a potential.  The reason is that the potential energy operator for 
particles in an external potential $V({\bf x})$ and a mutually interacting potential V({\bf x}-{\bf x}') is  
$\int d{\bf x}\xi^{\dagger}({\bf x})\xi({\bf x})V({\bf x})+\int d{\bf x}d{\bf x}'\xi^{\dagger}({\bf x})\xi({\bf x})\xi^{\dagger}({\bf x}')\xi({\bf x}') V({\bf x}-{\bf x}')$.  
This commutes with the non-relativistic collapse-generating operator $\sim\xi^{\dagger}({\bf x})\xi({\bf x})$ but does not 
commute with the relativistic collapse-generating operator  $\sim K^{1/2}\xi^{\dagger}({\bf x})K^{1/2}\xi({\bf x})$.  Thus, 
the energy increase  for relativistic particles will be modified from (\ref{8}), which is worth investigating\cite{sud}.

\section{Photon Number Decrease}

We shall now consider the effect of the collapse dynamics on a laser beam of finite length $\sim\sigma$, for example, a laser pulse.

We shall describe the initial state vector of the laser beam as the coherent state 
\begin{equation}\label{10}
|\psi,0\rangle\equiv e^{\beta\int d{\bf k}\alpha({\bf k})a^{\dagger}({\bf k})}|0\rangle e^{-\beta^{2}/2}, \hbox{ with } 
\alpha({\bf k})\equiv\frac{1}{(\pi/\sigma^{2})^{3/4}}e^{-({\bf k}-{\bf k}_{0})^{2}\sigma^{2}/2}.
\end{equation}
$\beta$ is a positive constant, whose square is the mean photon number, as we shall see below.

For simplicity, (\ref{10}) gives the width of the beam as $\sim\sigma$ also, where of course it is usually quite a bit smaller than the length: 
this has no consequence as the only relevant property employed is that the width, like the length, is many times larger than the wavelength.  

We define the state of $n$ photons as
\begin{equation}\label{11}
|n\rangle\equiv\frac{1}{\sqrt{n!}}\Bigg[\int d{\bf k}\alpha({\bf k})a^{\dagger}({\bf k})\Bigg]^{n}|0\rangle \hbox{ so } \langle n| m\rangle=\delta_{nm}.
\end{equation}
\noindent The initial density matrix is $\rho(0)=|\psi,0\rangle\langle \psi,0|$.  Thus, the probability that there are $n$ particles in the initial state is the Poisson distribution 
\begin{equation}\label{12}
 \langle n|\rho(0)|n\rangle=\frac{\beta^{2n}}{n!}e^{-\beta^{2}}\hbox { from which one finds that the initial mean number of photons is } \bar n(0)=\beta^{2}.
\end{equation}

It follows from Eq.(\ref{1}) for photons ($M=0$) that, to first order in $\lambda$, 
\begin{eqnarray}\label{13}
\langle n|\rho(t)|n\rangle&\approx&\frac{\beta^{2n}}{n!}e^{-\beta^{2}}
-\frac{\lambda t}{2M_{N}^{2}}\frac{1}{(2\pi)^{3}}(4\pi a^{2})^{3/2}\int d{\bf k}_{1}d{\bf k}_{2}d{\bf k}_{3}d{\bf k}_{4}
\sqrt{k_{1}k_{2}k_{3}k_{4}}                                   \nonumber\\
&&\cdot e^{-({\bf k}_{1}-{\bf k}_{2})^{2}a^{2}}\delta(-{\bf k}_{1}+{\bf k}_{2}-{\bf k}_{3}+{\bf k}_{4})\nonumber\\
&&\negmedspace\negmedspace\negmedspace\negmedspace\negmedspace\negmedspace\negmedspace\negmedspace
\negmedspace\negmedspace\negmedspace\negmedspace\negmedspace\negmedspace\negmedspace\negmedspace
\cdot\Big[\langle n|a^{\dagger}({\bf k}_{1})a({\bf k}_{2})a^{\dagger}({\bf k}_{3})a({\bf k}_{4})\rho(0)|n\rangle- 
 \langle n|a^{\dagger}({\bf k}_{3})a({\bf k}_{4})\rho(0)a^{\dagger}({\bf k}_{1})a({\bf k}_{2})|n\rangle+hc\Big].
\end{eqnarray}
\noindent All  terms are real so the bracketed terms in (\ref{13}) are equal to their Hermitian conjugate.   Note that the Hamiltonian term 
makes no contribution to this diagonal matrix element since $|n\rangle$ is very close to being an energy eigenstate,  
$H|n\rangle\approx nk_{0} |n\rangle$, so $\langle n|[H,  \rho(t)]|n\rangle\approx0$.

Since $k_{0}\sigma>>1$ is certainly true for a laser pulse, we can readily make the approximations   
$\alpha({\bf k}_{i})\sqrt{{\bf k}_{i}}\approx  \alpha({\bf k}_{i})\sqrt{{\bf k}_{0}}$, 
and  $\alpha^{2}({\bf k})\approx \delta ({\bf k}-{\bf k}_{0})$.

Using $a({\bf k})|\psi,0\rangle=\beta\alpha({\bf k})|\psi,0\rangle$, $a({\bf k})|n\rangle=\sqrt{n}\alpha({\bf k})|n-1\rangle$, 
$\langle n|\psi,0\rangle=\frac{\beta^{n}}{\sqrt{n!}}e^{-\beta^{2}/2}$,  and putting 
$[a({\bf k}_{2}),a^{\dagger}({\bf k}_{3}]=\delta({\bf k}_{2}-{\bf k}_{3})$ 
in the first bracketed term of (\ref{13}) we get:
\begin{eqnarray}\label{14}
\langle n|\rho(t)|n\rangle&\approx&\frac{\beta^{2n}}{n!}e^{-\beta^{2}}\Bigg[1
-\frac{\lambda t}{M_{N}^{2}}\Big(\frac{a^{2}}{\pi}\Big)^{3/2}\int d{\bf k}_{1}d{\bf k}_{2}d{\bf k}_{3}d{\bf k}_{4}
\sqrt{k_{1}k_{2}k_{3}k_{4}}                                    \nonumber\\
&&e^{-({\bf k}_{1}-{\bf k}_{2})^{2}a^{2}}\delta(-{\bf k}_{1}+{\bf k}_{2}-{\bf k}_{3}+{\bf k}_{4})\nonumber\\
&&\Big[\alpha({\bf k}_{1})\alpha({\bf k}_{2})\alpha({\bf k}_{3})\alpha({\bf k}_{4})[n(n-1)-n^{2}]+\alpha({\bf k}_{1})\alpha({\bf k}_{4})\delta({\bf k}_{2}-{\bf k}_{3})n\Big].
\end{eqnarray}

For the first bracketed term in (\ref{14}), upon setting  $\sqrt{k_{1}k_{2}k_{3}k_{4}} \approx k_{0}^{2}$, the integral may be performed: this is done in Appendix \ref{B}.  The result is  $k_{0}^{2}(2\pi/\sigma^{2})^{3/2}$, which is $<<$  the second term 
(see Eqs.(\ref{15a}, \ref{15b}) below) and so may be neglected.

So, it is the second bracketed term which is of interest. For the integral involving it, one may obtain a closed expression in the two limits $k_{0}a<<1$ and $k_{0}a>>1$:
\begin{subequations}
\begin{eqnarray}\label{15}
&&\frac{1}{(\pi/\sigma^{2})^{3/2}}\int d{\bf k}_{1}d{\bf k}_{2}k_{1}k_{2}e^{-({\bf k}_{1}-{\bf k}_{2})^{2}a^{2}}e^{-({\bf k}_{1}-{\bf k}_{0})^{2}\sigma^{2}}\nonumber\\
&\approx&\int d{\bf k}_{1}d{\bf k}_{2}k_{1}k_{2}e^{-({\bf k}_{1}-{\bf k}_{2})^{2}a^{2}}\delta({\bf k}_{1}-{\bf k}_{0})
=\int d{\bf k}_{2}k_{2}k_{0}e^{-({\bf k}_{2}-{\bf k}_{0})^{2}a^{2}}\nonumber\\
&\approx &  \int d{\bf k}_{2}k_{2}k_{0}e^{-k_{2}^{2}a^{2}} = k_{0}\frac{2\pi}{a^{4}}  \hbox{\qquad for $k_{0}a<<1$},\label{15a}\\
&\approx& \int d{\bf k}_{2}k_{2}k_{0}\Big(\frac{\pi}{a^{2}}\Big)^{3/2}\delta ({\bf k}_{2}-{\bf k}_{0})=k_{0}^{2}\Big(\frac{\pi}{a^{2}}\Big)^{3/2} \hbox{\qquad for $k_{0}a>>1$}.\label{15b}
\end{eqnarray}
\end{subequations}
\noindent 
Inserting Eqs.(\ref{15a}, \ref{15b}) into (\ref{14}), we obtain the results:
\begin{subequations}
\begin{eqnarray}\label{16}
\langle n|\rho(t)|n\rangle&\approx& 
\frac{\beta^{2n}}{n!}e^{-\beta^{2}}\Bigg[1-4\pi^{1/2}n\lambda t\frac{\lambdabar_{N}^{2}}{\lambda_{0}a}\Bigg]     \hbox{\qquad for $k_{0}a<<1$},\label{16a}\\
\langle n|\rho(t)|n\rangle&\approx&\frac{\beta^{2n}}{n!}e^{-\beta^{2}}\Bigg[1-n\lambda t\Bigg(\frac{\lambdabar_{N}}{\lambda_{0}} \Bigg)^{2}\Bigg] \hbox{\qquad for $k_{0}a>>1$}.
\label{16b}
\end{eqnarray}
\end{subequations}
\noindent  Setting $\beta^{2}=\bar n(0)$ and approximating $\beta^{2}(\beta^{2} +1)\approx \beta^{4}$, the mean number of photons in such a pulse is calculated to decrease as 
\begin{subequations}
 \begin{eqnarray}\label{17}
\bar n(t)&=&\sum_{n=0}^{\infty}n\langle n|\rho(t)|n\rangle\nonumber\\
&\approx& \bar n(0)\Bigg[1-4\pi^{1/2}\bar n (0)\lambda t\frac{\lambdabar_{N}^{2}}{\lambda_{0}a}\Bigg]   \hbox{\qquad for $k_{0}a<<1$}        \label{17a}       \\
&\approx& \bar n(0)\Bigg[1-\bar n (0)\lambda t\Bigg(\frac{\lambdabar_{N}}{\lambda_{0}} \Bigg)^{2}\Bigg] \hbox{\qquad for $k_{0}a>>1$}.\label{17b}  
\end{eqnarray}
\end{subequations}

Eqs.(\ref{17a}, \ref{17b}) are the result we have been seeking. 
We see  that the rate of photon loss is largest for a large number of photons in a pulse or for a small photon wavelength.  
Let us consider experimental situations where these dependencies come to the fore. 

Considering the case of a large number of photons, at the Vulcan laser facility\cite{mich} there are presently generated high intensity laser beam pulses  containing  energy $E_{p}\approx 500$J, 
although  with a fairly large wavelength, $\lambda_{0}= 1053$nm, in the infrared. There are then 
$\bar n(0)= E_{p}/(hc/\lambda_{0})\approx 2.5\times10^{21}$ photons in a pulse (pulse length $\sigma \approx .1$mm).  In this case $k_{0}a\approx .6$ lies between the validity  regions of  (\ref{17a}) or (\ref{17b}), but (\ref{17a}) gives $\bar n(t)\approx2.5\times10^{21}[1-.75\times10^{4}\lambda t]$ and (\ref{17b}) gives 
$\bar n(t)\approx2.5\times10^{21}[1-10^{4}\lambda t]$
                 
Considering the case of energetic photons, the most intense, XFEL (X-ray Free Electron Laser)  pulses provide the attendant 
increase of the $(\lambdabar_{N}/\lambda_{0})^{2}$ factor in Eqs.(\ref{17a}, \ref{17b}), although there is a smaller  $\bar n(0)$.  The LCLSII (Stanford  Linear Coherent Light Source)\cite{LCLS} specifies its laser pulses as 
containing $\approx 10^{12}$ photons, each of 8.3KeV ($\approx 1$mJ/pulse, pulse length $\sigma\approx .15$mm). Then, 
$\lambdabar_{N}/\lambda_{0}\approx 10^{-13}\hbox{cm}/10^{-8}\hbox{cm}=10^{-5}$. With these values, (\ref{17b}) becomes
\begin{equation}\label{18}
\bar n(t)=10^{12}[1-100\lambda t].
\end{equation}

In both these cases, one might at least imagine an experiment measuring the loss of photons from a pulse with the pulse bouncing  
back and forth between mirrors many times to be accessible over, say, 1s. This would have to 
 contend with competing loss mechanisms such as attendant loss at each bounce, scattering losses from the 
gas between the mirrors.  Supposing these effects could be compensated for, and the accuracy of the measurement was 1\% with no loss observed, 
this would place a limit $\lambda\lesssim10^{-4}-10^{-6}$s$^{-1}$: the present best upper limit\cite{Bassi} is around $\lambda\leq 10^{-9}-10^{-10}$s$^{-1}$.

\section{ Photon excitation.}
The operators in the density matrix evolution equation do not change the number of photons.  To first order in $\lambda$, there is the probability of conversion of a photon of momentum 
${\bf k}_{0}$ to one of momentum ${\bf k}$.  We shall indeed see that this compensates the resulting loss of photons from the beam presented in Sec. III,  for which $Tr\rho(t)<1$.  Thus, for the combined processes of photon loss 
and photon excitation, $Tr\rho(t)=1$.  

We shall also see that the energy increase in Sec. II, Eq.(\ref{6}), is explained, to first order in $\lambda$, by replacement of a photon of energy 
$k_{0}$ by one of energy $k$. 

Then we shall consider a consequence of the predicted excited photon distribution. 

We need a complete set of orthogonal one-photon states, of which one state is the photon state in the laser beam, $\int d{\bf k} \alpha({\bf k}) a^{\dagger}({\bf k})|0\rangle$.  Since we have chosen 
$\alpha({\bf k})$ to have the form of the ground state of a three dimensional harmonic oscillator in the variable ${\bf k}-{\bf k}_{0}$, the orthogonal set is 
readily supplied as $\mu_{{\bf s}}^{\dagger}|0\rangle\equiv\int d{\bf k} \chi_{\bf s}({\bf k}) a^{\dagger}({\bf k} )|0\rangle$.  Here $\chi_{{\bf s}}({\bf k})\equiv N_{{\bf s}}H_{s_{1}}H_{s_{2}}H_{s_{3}}\alpha({\bf k})$, 
where $H_{s_{i}}$ is a Hermite polynomials in  $ k_{i}-k_{0i}$, and the three indices ${\bf s}\equiv (s_{1},s_{2},s_{3})$ take on all integer values $\geq0$ 
(so  $\chi_{0,0,0}({\bf k}) =\alpha ({\bf k})$).

We wish to consider the expectation value of the density matrix for an $n+1$ particle state (i.e., the probability that this state is occupied), where $n$ particles comprise the state $|n\rangle$ and one more particle is ``almost" in the momentum eigenstate $|{\bf k}\rangle=a^{\dagger}({\bf k})|0\rangle$. By ``almost" is meant that the state is orthogonal to 
$\mu_{0,0,0}^{\dagger}|0\rangle$.  To this end, we define 
the projection operator $P\equiv 1-\mu_{0,0,0}^{\dagger}|0\rangle\langle 0|\mu_{0,0,0}$.  We shall also find it useful to define 
$\gamma_{n}^{\dagger}\equiv \frac{1}{\sqrt{n!}}[\int d{\bf k}\alpha({\bf k})a^{\dagger}({\bf k})]^{n}$ so $\gamma_{n}^{\dagger}|0\rangle=|n\rangle$.  Thus, the $n+1$ particle state 
is $\gamma_{n}^{\dagger}P|{\bf k}\rangle$.  

Since $1=\sum_{{\bf s}}\mu_{{\bf s}}^{\dagger}|0\rangle\langle 0|\mu_{{\bf s}}$, it follows that $|{\bf k}\rangle=\sum_{{\bf s}}\chi_{{\bf s}}({\bf k})\mu_{{\bf s}}^{\dagger}|0\rangle$ and 
$P|{\bf k}\rangle=\sum_{{\bf s}\neq(0,0,0)}\chi_{{\bf s}}({\bf k})\mu_{{\bf s}}^{\dagger}|0\rangle$ so $\langle 0|\gamma_{1} P|{\bf k}\rangle=0$. Therefore,  
$\langle m|\gamma_{n}^{\dagger}P|{\bf k}\rangle=\delta_{m, n+1}\frac{1}{\sqrt{n+1}} \langle 0|\gamma_{1} P|{\bf k}\rangle=0$, and so  $\rho(0)\gamma_{n}^{\dagger}P|{\bf k}\rangle=0$.
          
  Then, the density matrix diagonal element is, to first order 
in $\lambda$, 
\begin{eqnarray}\label{19}
\langle{\bf k}|P\gamma_{n}\rho(t)\gamma_{n}^{\dagger} P|{\bf k}\rangle&=&
\frac{\lambda t}{2M_{N}^{2}}\Big(\frac{a^{2}}{\pi}\Big)^{3/2}\int d{\bf k}_{1}d{\bf k}_{2}d{\bf k}_{3}d{\bf k}_{4}
\sqrt{k_{1}k_{2}k_{3}k_{4}}                                
 e^{-({\bf k}_{1}-{\bf k}_{2})^{2}a^{2}}\delta(-{\bf k}_{1}+{\bf k}_{2}-{\bf k}_{3}+{\bf k}_{4})\nonumber\\
&&
\cdot
\sum_{{\bf s}, {\bf s}'\neq(0,0,0)}\chi_{{\bf s}}({\bf k}) \langle 0|\mu_{\bf s}\gamma_{n}a^{\dagger}({\bf k}_{3})a({\bf k}_{4})\rho(0)a^{\dagger}({\bf k}_{1})a({\bf k}_{2})
\gamma_{n}^{\dagger}\mu_{s'}^{\dagger}|0\rangle\chi_{{\bf s}'}({\bf k})+hc\Big]\nonumber\\
&=& \frac{\lambda t}{M_{N}^{2}}\Big(\frac{a^{2}}{\pi}\Big)^{3/2}\int d{\bf k}_{1}d{\bf k}_{2}d{\bf k}_{3}d{\bf k}_{4}
\sqrt{k_{1}k_{2}k_{3}k_{4}}                                
 e^{-({\bf k}_{1}-{\bf k}_{2})^{2}a^{2}}\delta(-{\bf k}_{1}+{\bf k}_{2}-{\bf k}_{3}+{\bf k}_{4})\nonumber\\
&&
\cdot
\beta^{2}\frac{\beta^{2n}}{n!}e^{-\beta^{2}}\sum_{{\bf s}, {\bf s}'\neq(0,0,0)}\chi_{{\bf s}}({\bf k}_{3})\chi_{{\bf s}}({\bf k})\chi_{{\bf s}'}({\bf k})\chi_{{\bf s}'}({\bf k}_{2})\alpha({\bf k}_{1})\alpha({\bf k}_{4})
\nonumber\\
&=& \frac{\lambda t}{M_{N}^{2}}\Big(\frac{a^{2}}{\pi}\Big)^{3/2}\int d{\bf k}_{1}d{\bf k}_{2}d{\bf k}_{3}d{\bf k}_{4}
\sqrt{k_{1}k_{2}k_{3}k_{4}}                                
 e^{-({\bf k}_{1}-{\bf k}_{2})^{2}a^{2}}\delta(-{\bf k}_{1}+{\bf k}_{2}-{\bf k}_{3}+{\bf k}_{4})\nonumber\\
&&
\cdot
\beta^{2}\frac{\beta^{2n}}{n!}e^{-\beta^{2}}[\delta({\bf k}-{\bf k}_{3})- \alpha({\bf k}) \alpha({\bf k}_{3})][\delta({\bf k}-{\bf k}_{2})- \alpha({\bf k}) \alpha({\bf k}_{2})]\alpha({\bf k}_{1})\alpha({\bf k}_{4}).
\end{eqnarray}
\noindent (First, the matrix elements were evaluated. Second, the completeness relation $\sum_{{\bf s}\neq (0,0,0)}\chi_{{\bf s}}({\bf k})\chi_{{\bf s}}({\bf k}') + \alpha({\bf k})\alpha({\bf k}')
=\delta ({\bf k}-{\bf k}')$ was employed.)

It is the term $\sim\delta({\bf k}-{\bf k}_{3})\delta({\bf k}-{\bf k}_{2})$ that provides the important contribution. 
Summing (\ref{19}) over all $n$ (setting $\beta^{2}=\bar n(0)$) gives the probability density for the presence of a photon of momentum ${\bf k}$:
\begin{eqnarray}\label{20}
{\cal P}({\bf k})&=& \bar n(0)\frac{\lambda t}{M_{N}^{2}}\Big(\frac{a^{2}}{\pi}\Big)^{3/2}\int d{\bf k}_{1}k_{1}k                              
 e^{-({\bf k}_{1}-{\bf k})^{2}a^{2}}\Big(\frac{\sigma^{2}}{\pi}\Big)^{3/2} e^{-({\bf k}_{1}-{\bf k}_{0})^{2}\sigma^{2}}+R({\bf k})
\end{eqnarray}
\noindent with the contribution of the other terms being 
\begin{eqnarray}\label{21}
R({\bf k})&\equiv&\bar n(0)\frac{\lambda t}{M_{N}^{2}}\Big(\frac{a^{2}}{\pi}\Big)^{3/2}\int d{\bf k}_{1}d{\bf k}_{2}d{\bf k}_{3}d{\bf k}_{4}\sqrt{k_{1}k_{2}k_{3}k_{4}}\delta(-{\bf k}_{1}+{\bf k}_{2}-{\bf k}_{3}+{\bf k}_{4}) e^{-({\bf k}_{1}-{\bf k}_{2})^{2}a^{2}}\nonumber\\
&&\alpha({\bf k}_{1})\alpha({\bf k}_{2})\alpha({\bf k}_{3})\alpha({\bf k}_{4})\Bigg[\alpha^{2}({\bf k})-2\delta({\bf k}_{2}-{\bf k}))\Bigg]\nonumber\\
&\approx&\bar n(0)\frac{\lambda t}{M_{N}^{2}}\Big(\frac{a^{2}}{\pi}\Big)^{3/2}\Big[k_{0}^{2}2^{3/2}e^{-({\bf k}-{\bf k}_{0})^{2}\sigma^{2}}-2k_{0}^{2}\Big(\frac{4}{3}\Big)^{3/2} 
e^{-({\bf k}-{\bf k}_{0})^{2}2\sigma^{2}/3}\Big],
\end{eqnarray}
\noindent where the result in the last line of (\ref{21}) is obtained in Appendix \ref{B}.  We can neglect $R({\bf k})$ with respect to the first term.  For, 
(\ref{21}) only makes a contribution  for ${\bf k}={\bf k}_{0}+o(1/\sigma)$, which cannot be distinguished from  
a photon in the laser pulse. Moreover, 
we note that the integrated probability  contribution of the term in the bracket in (\ref{21}) is $-k_{0}^{2}(2\pi/\sigma^{2})^{3/2}$ which precisely cancels (\ref{B4}), the term dropped from the photon loss expression (\ref{14}) because it is 
 smaller than the term kept.  
 
Employing the approximation $\Big(\frac{\sigma^{2}}{\pi}\Big)^{3/2} e^{-({\bf k}_{1}-{\bf k}_{0})^{2}\sigma^{2}}\approx \delta ({\bf k}_{1}-{\bf k}_{0})$ in (\ref{20}), the result we have been seeking,  the probability density of the existence of collapse-excited photons, is:
\begin{equation}\label{22}
{\cal P}({\bf k})=\bar n(0)\frac{\lambda t}{M_{N}^{2}}\Big(\frac{a^{2}}{\pi}\Big)^{3/2}k_{0}k                              
 e^{-({\bf k}_{0}-{\bf k})^{2}a^{2}}.  
\end{equation}

First, to connect with the result of Section III, to verify that the loss of photons there and the gain (\ref{22}) here  account for all photons.    The trace of the density matrix over the laser beam states is found from Eq.(\ref{14})'s second bracketed term (with the integral involved replaced by the last term in the second line of (\ref{15a})):
\begin{equation}\label{23}
\sum_{n}\langle n|\rho(t)|n\rangle\approx\Bigg[1
-\bar n(0)\frac{\lambda t}{M_{N}^{2}}\Big(\frac{a^{2}}{\pi}\Big)^{3/2}\int d{\bf k}_{1}k_{1}k_{0}e^{-({\bf k}_{1}-{\bf k}_{0})^{2}a^{2}}\Bigg].
\end{equation}
When we add this to the trace over the excited photon states, $\int d{\bf k}{\cal P}({\bf k})$,  we obtain the correct result that the trace over all states is 1 to order $\lambda t$.

Second, to connect with the result of Section II, to verify that result gives the energy increase in this instance.  ${\cal P}({\bf k})/\bar n(0)$  is the probability that a single photon has been converted to momentum ${\bf k}$ from 
momentum ${\bf k}_{0}$.
The energy change for such a photon is therefore $k-k_{0}$ and  for  $\bar n(0)$ photons is therefore, using (\ref{22}),
\begin{equation}\label{24}
\bar E(t)-\bar E(0)=\bar n(0)\frac{\lambda t}{M_{N}^{2}}\Big(\frac{a^{2}}{\pi}\Big)^{3/2}\int d{\bf k}(k-k_{0})k_{0}k                              
 e^{-({\bf k}_{0}-{\bf k})^{2}a^{2}}. 
\end{equation}
On the other hand, Eq. (\ref{6}) specialized to photons is
\begin{eqnarray}\label{25}
\frac{\partial}{\partial t}\bar H(t)&=&\frac{\lambda}{M_{N}^{2}}\Big(\frac{a^{2}}{\pi}\Big)^{3/2}Tr\rho(t)\int d{\bf k}_{1} a^{\dagger}({\bf k}_{1})a({\bf k}_{1})
\int d{\bf k}_{2}k_{1}k_{2}(k_{2}-k_{1})
e^{-({\bf k}_{2}-{\bf k}_{1})^{2}a^{2}}.
\end{eqnarray}
\noindent To first order in $\lambda$, $\rho(t)\rightarrow \rho(0)$ and, since $\int d{\bf k}_{1} a^{\dagger}({\bf k}_{1})a({\bf k}_{1})$ is the beam photon number operator, the trace is 
$\bar n(0)$ and we see that (\ref{25}) is  then identical to  (\ref{24}).

Now, let's turn to an application of  (\ref{22}).  Among the most energetic of CW lasers is the carbon dioxide laser, with a wavelength 
$\lambda_{0}\approx 1000$nm, with a few hundred kw beam achieved\cite{CWlaser}.  We shall consider  a 1 megawatt beam which has been suggested achieved for 
military purposes.

Since $\lambda_{0}<<a$, the probability density distribution  (\ref{22}) is well approximated by 
\begin{equation}\label{26}
{\cal P}({\bf k})\approx\bar n(0)\frac{\lambda t}{M_{N}^{2}}\Big(\frac{a^{2}}{\pi}\Big)^{3/2}k_{0}k                              
 e^{-k^{2}a^{2}}.  
\end{equation}
\noindent Thus, each photon in the beam has a probability of being converted to an excited photon of some wavelength characterized by the scale $\lesssim2\pi a$, 
and the  photons sprayed out have a spherically symmetric distribution. 

 So, suppose we consider an experiment continuously monitoring a 3m length of such a beam (traveling from the laser to some kind of absorber), looking for 
sprayed photons emerging from the beam in the suggested energy range.

The probability of any such photon appearing is 
\begin{equation}\label{27}
\Gamma\bar n(0)t\equiv\int d{\bf k}{\cal P}({\bf k})\approx\bar n(0)\frac{\lambda t\hbar^{2}}{M_{N}^{2}c^{2}}\Big(\frac{a^{2}}{\pi}\Big)^{3/2}k_{0}\frac{2\pi}{a^{4}}= 
4\pi^{1/2}\bar n(0)\lambda t \frac{\lambdabar_{N}^{2}}{\lambda_{0}a}.                         
\end{equation}
Although our calculation has been to only first order in $\lambda$, and therefore only holds for small values of the probability, 
if we regard $\Gamma$ defined above as a time-translation-invariant rate of production of anomalous photons per laser photon, we may 
allow $\Gamma\bar n(0)t$, the number of photons produced in time $t$, to exceed 1.

The energy in a 3m beam length of a 1 megawatt beam is $3\times10^{6}/c=10^{-2}J$.  The energy in a single photon is $hc/\lambda_{0}\approx 2\times10^{-19}J$. 
Thus there are $\bar n(0)\approx5\times 10^{16}$photons in that length and so,  from   (\ref{27}),    $\Gamma\bar n(0)t\approx .14\lambda t$.  In one year$\approx 3\times 10^{7}$s, 
according to  Eq.(\ref{27}), one expects $\approx 4\times10^{6}\lambda$ anomalous ``sprayed" photons.  If no photons are seen, and there is 5\% probability of experimental error, 
that would place a limit $\lambda\lesssim10^{-8}$s$^{-1}$.    
 
 \section{Effect on Cosmic Blackbody Radiation}
 
 We wish to examine how the distribution of blackbody photons is altered by the collapse excitation mechanism, and apply the result to 
 the cosmic blackbody spectrum.  
 
 \subsection{Effect of Collapse on Blackbody Radiation}
Blackbody radiation is described by the thermal density matrix $\rho(0)=e^{-\beta H}/Tre^{-\beta H}$, where $\beta\equiv 1/k_{B}T$ and $H=\int d{\bf k}ka^{\dagger}({\bf k})a^{\dagger}({\bf k})$. 
It is traditional to work with box-normalized momenta rather than with continuous momenta, so we shall do that,  writing ${\bf k}_{i}= 
(2\pi/L){\bf n}_{i}=(2\pi/L)(n_{ix}, n_{iy},n_{iz})$ with the $ n_{ij}$ as integers $-\infty< n_{ij}<\infty$. Likewise, we 
write  $\int d{\bf k}_{{\bf i}}=(2\pi/L)^{3}\sum_{{\bf n}_{i}}$, and  $a({\bf k}_{i}) =(L/2\pi)^{3/2}a_{{\bf i}} $,  
$\delta ( {\bf k}_{{\bf i}}-{\bf k}_{{\bf i}'} )=(L/2\pi)^{3}\delta_{{\bf n}_{i},{\bf n}_{i}'}$, so $[a_{i},a^{\dagger}_{i'}]= \delta_{{\bf n}_{i},{\bf n}_{i}'}$
and $H=\sum_{m}k_{m}a^{\dagger}_{m}a_{m}$.  

We may now write the expression for the time rate of change of the mean energy in a single mode, $\bar \epsilon_{s}\equiv k_{s}a^{\dagger}_{s}a_{s}$ to 
first order in $\lambda$.  This is essentially Eq.(\ref{4}) written in terms of box-normalized momenta, but without summing over all modes, and with an extra factor of 2 
because there are two polarizations:
\begin{eqnarray}\label{28}
\frac{\partial}{\partial t}\bar \epsilon_{s}&=&-\frac{\lambda}{M_{N}^{2}}\Big(\frac{a^{2}}{\pi}\Big)^{3/2}Tr\rho(0)\Big(\frac{2\pi}{L}\Big)^{3}
\sum_{{\bf n}_{1}}k_{s}
\sqrt{k_{1},k_{2},k_{3},k_{4}}k_{s}e^{-({\bf k}_{1}-{\bf k}_{2})^{2}a^{2}}\delta _{-n_{1}+n_{2}-n_{3}+n_{4}}[a^{\dagger}_{1}a_{2},[a^{\dagger}_{3}a_{4},a^{\dagger}_{s}a_{s}]]\nonumber\\
&=&-\frac{2\lambda}{M_{N}^{2}}\Big(\frac{a^{2}}{\pi}\Big)^{3/2}Tr\rho(0)\Big(\frac{2\pi}{L}\Big)^{3}k_{s}^{2}\sum_{{\bf n}_{1}}
k_{1}[a^{\dagger}_{s}a_{s}-a^{\dagger}_{1}a_{1}]e^{-({\bf k}_{s}-{\bf k}_{1})^{2}a^{2}}.
\end{eqnarray}
 \noindent As a check, we confirm that the mean number of photons is not changed by the collapse process, as follows. 
 We note that if $\epsilon_{{\bf s}}$ is replaced by $n_{{\bf s}}=\epsilon_{{\bf s}}/k_{s}$ and all modes are summed over, 
 the left side of (\ref{28}) is $\frac{\partial}{\partial t}\bar n$, and with one less factor of $ k_{s}$, the sum over ${\bf n}_{s}$ of the right side of (\ref{28}) 
 vanishes. 
 
 We now may take the trace in Eq.(\ref{28}).  Since
 \[
 Tre^{-\beta k_{m}a^{\dagger}_{m}a_{m}}=\sum_{j=0}^{\infty}\langle 0| \frac{a_{m}^{j}}{\sqrt{j!}} e^{-\beta k_{m}a^{\dagger}_{m}a_{m}} \frac{a_{m}^{\dagger j}}{\sqrt{j!}}|0\rangle
 = \sum_{j=0}^{\infty}  e^{-\beta k_{m}j}                   =\frac{1}{1-e^{-\beta k_{m}}}
 \]
\noindent and so  
\[Tr\rho(0)a^{\dagger}_{m}a_{m}=[1-e^{-\beta k_{m}}]\sum_{j=0} ^{\infty}e^{-\beta k_{m}j}j=\frac{1}{e^{\beta k_{m}}-1},
\]
\noindent we obtain 
\begin{eqnarray}\label{29}
\bar \epsilon_{s}(t)&=&\frac{2k_{s}}{e^{\beta k_{s}}-1}-\frac{2\lambda t}{M_{N}^{2}}\Big(\frac{a^{2}}{\pi}\Big)^{3/2}\Big(\frac{2\pi}{L}\Big)^{3}k_{s}^{2}\sum_{{\bf n}_{1}}
k_{1}\Bigg[ \frac{1}{e^{\beta k_{s}}-1}-\frac{1}{e^{\beta k_{1}}-1}\Big]e^{-({\bf k}_{s}-{\bf k}_{1})^{2}a^{2}}.\nonumber\\
&=&\frac{2k_{s}}{e^{\beta k_{s}}-1}\Bigg[1-\frac{\lambda t}{M_{N}^{2}}\Big(\frac{a^{2}}{\pi}\Big)^{3/2}k_{s}\int d{\bf k}_{1}k_{1}e^{-({\bf k}_{s}-{\bf k}_{1})^{2}a^{2}}\Bigg]\nonumber\\
&&\qquad\qquad\qquad+\frac{2\lambda t}{M_{N}^{2}}\Big(\frac{a^{2}}{\pi}\Big)^{3/2}k_{s}^{2}\int d{\bf k}_{1}k_{1}e^{-({\bf k}_{s}-{\bf k}_{1})^{2}a^{2}}\frac{1}{e^{\beta k_{1}}-1}
\end{eqnarray}
\noindent where we have replaced the sum by an integral, returning to the continuum momentum for that variable, but so far keeping the discrete momentum for $k_{s}$.

According to the first line of Eq.(\ref{29}), as time progresses, photons are kicked out of the mode with momentum $k_{s}$ According to the second line, photons are also kicked into this mode from all the other modes. 
As we shall see below, loss from high probability modes is the rule, since the photons lost are most likely kicked to higher energy, low probability modes, 
characterized by $k\sim 1/a$.

(As a final check, we note that the total mean energy increase calculated in Section II agrees with the result here.  If we apply Eq.(\ref{4}) to first order in $\lambda$, 
where $a^{\dagger}({\bf k}_{1})a({\bf k}_{1})$ is replaced by $(L/2\pi)^{3}a_{1}^{\dagger}a_{1}$, so that the trace equation delineated above applies, with 
the extra factor of 2 for the two polarizations, the result is
\begin{equation}\label{30}
\bar E(t)=\Big(\frac{L}{2\pi}\Big)^{3}\Bigg[ \int d{\bf k} \frac{2k}{e^{\beta k}-1} +   
\frac{2\lambda}{M_{N}^{2}}\Big(\frac{a^{2}}{\pi}\Big)^{3/2}\int d{\bf k}_{1}d{\bf k}_{2}k_{1}k_{2}(k_{2}-k_{1})\frac{1}{e^{\beta k_{1}}-1}e^{-({\bf k}_{1}-{\bf k}_{2})^{2}a^{2}}\Bigg].
\end{equation}
\noindent This is the same as the integral of Eq.(\ref{29}), with appropriate renaming of integration variables.)

\subsection{Effect of Collapse on Cosmic Microwave Radiation}

We wish to use (\ref{29}) to calculate the change, due to the collapse process, in photon number in the modes of wavelength $\approx .05$cm to $50$cm in the cosmological microwave radiation we receive, as the radiation travels toward us.  

The distance of the radiation at any time $t$ after recombination is characterized by the redshift parameter $Z(t)$. 
Taking $t=0$ as the time of recombination, which occurred $\approx 400, 000$yr after the big bang, it is found that $Z(0)\approx 1000$.  Denoting by a subscript 0 the present value of a quantity, the  radiation temperature T(t) is related to the present radiation temperature $T_{0}\approx 2.7^{\circ}K$ by the relation $T(t)=(1+Z(t))T_{0}$.   Thus, at recombination, the temperature was  $T(0)\approx 1000T_{0}\approx 3000^{\circ}K$.   

If we  neglect the time differences between the Hubble time $\approx 14\times 10^{9}$yr, the age of the universe, and 
the time since recombination $t_{0}$, 
the time evolution of $Z$, according to Hubble's law,  is $Z(t)\approx1000[1-\frac{t}{t_{0}}]$. Any length $\ell$, such as a wavelength $\lambda$ or the length $L$ of the side of the normalization box 
behaves as $\ell(t)=\ell_{0}/(1+Z(t))$.    

    We shall simplify (\ref{29}) by the approximation in the integrals $e^{-({\bf k}_{s}-{\bf k}_{1})^{2}a^{2}}\approx e^{- k_{1}^{2}a^{2}}$.  This is certainly valid at present since 
    the high probabiity modes' wavelengths are much larger than $a$. However, we must consider this approximation at all times. 
We note that $(k_{s}-k_{1})a=(k_{0s}-k_{01})(1+Z)a$, and $(1+Z)a$ ranges from its its value at recombination, $10^{-2}$cm, to its present value, $10^{-5}$cm.  For this approximation to be valid, 
then $k_{0s}<<[(1+Z)a]^{-1}$  for the full range of $Z$.  This certainly is not true for the shorter wavelengths at recombination time, e.g., for $\lambda_{0s}\approx .05$cm, then  $k_{0s}\approx 100$ is equal to 
$[(1+Z)a]^{-1}$.  So, we may just take it as surely valid for longer wavelengths, say $\lambda_{0}\gtrsim.5$ cm, or consider that, since  it is valid for at least part of the photon journey, the result may be at least approximately applied to shorter wavelengths.  

With this approximation we may immediately evaluate the integral in the bracket in Eq.(\ref{29}), $\int d{\bf k}_{1}k_{1}e^{- k_{1}^{2}a^{2}}=2\pi/a^{4}$.  
   
The other integral in  Eq.(\ref{29}), is approximated as 
\[
\int d{\bf k}_{1}k_{1}e^{- k_{1}^{2}a^{2}}\frac{1}{e^{\beta k_{1}}-1}\approx \int d{\bf k}_{1}k_{1}\frac{1}{e^{\beta k_{1}}-1}=4\pi \Gamma(4)\zeta(4)/\beta^{4}\approx24\pi/\beta^{4}=12(2\pi)^{5}/\lambda_{Th}^{4}
\]
\noindent where we write $2\pi\beta=hc/k_{B}T\equiv\lambda_{Th}$ is the thermal wavelength, with $\lambda_{Th0}\approx .5$cm.
Here we have made the approximation $[e^{\beta k_{1}}-1]^{-1} e^{- k_{1}^{2}a^{2}}\approx [e^{\beta k_{1}}-1] $, i.e., we may set 
$e^{- k_{1}^{2}a^{2}}\approx1$.  To see this, first note that $\beta k=\beta_{0}k_{0}= \lambda_{Th}/\lambda_{0}$.  
Thus, $e^{-\beta k_{1}}e^{- k_{1}^{2}a^{2}}\approx e^{-.1 k_{10}}e^{- (k_{10}Z10^{-5})^{2}}$, and so $e^{-.1 k_{10}}$ dominates the integral for the full range of $Z$.
We have also approximated the Riemann 
zeta function $\zeta(4)=1.08...\approx 1$. 

Therefore, (\ref{29}) becomes
\begin{equation}\label{31}
\bar \epsilon_{s}(t)=\frac{2k_{s}}{e^{\beta k_{s}}-1}\Bigg[1-4\pi^{1/2}\lambda t\frac{\lambdabar_{N}^{2}}{a\lambda_{s}}\Bigg]+
k_{s}\frac{24(2\pi)^{6}}{\pi^{3/2}}\lambda t\frac{\lambdabar_{N}^{2}a^{3}}{\lambda_{Th}^{4}\lambda_{s}}.
\end{equation}
\noindent We discard the gain (last) term of (\ref{31}) as loss of photons massively predominates: the ratio of the loss term (second term in the bracket $\times 2k_{s}$) to the gain term is $\approx 3\times 10^{14}$ (it is dominated by $a^{-4}$). 

We want the number of photons in each mode, not the energy, so we divide by $k_{s} $.  Also, we are interested in the number/volume  in any mode with the same frequency. 
so we multiply the right hand side of (\ref{31}) by 
\[
1=\Big(\frac{L}{2\pi}\Big)^{3}\int_{\Omega}d{\bf k}_{s}= V\frac{1}{(2\pi)^{3}}4\pi k^{2}dk=V\frac{4\pi}{c^{3}}\nu^{2}d\nu 
\]
where, since we are returning to continuum variables, we replace $ k_{s}$ by $k$, we denote by $V$ the (normalization) volume containing the radiation, and 
then have converted  from variable $k$ to frequency $\nu$. Denoting by $\bar n(\nu, t)d\nu$ the number of photons with energy between $h\nu$ and $h(\nu+d\nu)$ in the volume $V$, 
we therefore have 
\begin{subequations}
\begin{eqnarray}
\bar n(\nu, t)d\nu&=&V\frac{8\pi\nu^{2}d\nu}{c^{3}[e^{h\nu/k_{B}T}-1]}\Bigg[1-4\pi^{1/2}\lambda t\nu\frac{\lambdabar_{N}^{2}}{ac}\Bigg],\label{32a}\\
\frac{d}{dt}\bar n(\nu, t)d\nu&=&-V\frac{32\pi^{3/2}\nu^{3}d\nu}{c^{3} [e^{h\nu/k_{B}T}-1]}                      \lambda \frac{\lambdabar_{N}^{2}}{ac}.\label{32b}
\end{eqnarray}
\end{subequations}

We have taken the time derivative in (\ref{32b}) to obtain the expression for the rate of photon loss over a short time interval.  To get the total photon loss 
over $t_{0}$, we first express (\ref{32b}) in terms of present variables, and then integrate over $t$.  We note that $V(t)=V_{0}/[1+Z(t)]^{3}$,  
$\nu(t)=\nu_{0}[1+Z(t)]^{3}$ and $h\nu(t)/k_{B}T(t)=h\nu_{0}/k_{B}T_{0}$.  Famously, the no-loss blackbody spectrum (the factor multiplying the 
bracket in (\ref{32a})) is time-independent, but the loss term has an extra $\nu$ factor, and so acquires an extra $1+Z(t)$ factor:
\begin{eqnarray}\label{33}
    \bar n(\nu_{0}, t_{0})/V_{0}&=&\frac{8\pi\nu_{0}^{2}}{c^{3}[e^{h\nu_{0}/k_{B}T_{0}}-1]}\Bigg[1-4\pi^{1/2} \nu_{0}\frac{\lambdabar_{N}^{2}}{ac}\lambda\int_{0}^{t_{0}}dt[1+Z(t)] \Bigg]\nonumber\\
    &=&\frac{8\pi\nu_{0}^{2}}{c^{3}[e^{h\nu_{0}/k_{B}T_{0}}-1]}\Bigg[1-4\pi^{1/2} \nu_{0}\frac{\lambdabar_{N}^{2}}{ac}500\lambda t_{0} \Bigg]\nonumber\\
    &\approx&\frac{8\pi\nu_{0}^{2}}{c^{3}[e^{h\nu_{0}/k_{B}T_{0}}-1]}\Bigg[1-.6\frac{\lambda}{(\lambda_{0}/.1)} \Bigg].
\end{eqnarray}
\noindent In the last line of (\ref{33}), the unit of $\lambda$ is sec$^{-1}$ and $\lambda_{0}$ is cm: we note that the peak of the energy per unit wavelength spectrum  is $\approx .1$cm.
In going from the second to the third line, we have used $t_{0}\approx40\times10^{16}$s and $\lambdabar_{N}\approx .5$cm. 

Eq.(\ref{31}) is the result we have sought. We see that the spectrum of the radiation received is altered, in that over its travel to us, collapse reduces the mean number of photons/volume  in each mode by an amount inversely proportional to the wavelength.  

The cosmic microwave radiation is experimentally found to have the blackbody form (excepting the famous anisotropies).  The 
temperature of the radiation is quoted as\cite{Fixsen} $2.72548\pm .00057^{\circ}K$, so error/temperature $\Delta\approx2\times 10^{-4}$.  
If the temperature of the pure blackbody spectrum is $T_{0}[1-\Delta]$, it may be written as \ignorespaces
\begin{eqnarray}\label{34}
      \bar n(\nu_{0}, t_{0})/V_{0}&=&\frac{8\pi\nu_{0}^{2}}{c^{3}[e^{h\nu_{0}/k_{B}T_{0}[1-\Delta]}-1]} 
\approx\frac{8\pi\nu_{0}^{2}}{c^{3}[e^{h\nu_{0}/k_{B}T_{0}}-1]} \Bigg[1- \frac{e^{\lambda_{Tho}/\lambda_{0}}(\lambda_{Tho}/\lambda_{0})\Delta}{[e^{\lambda_{Tho}/\lambda_{0}}-1]}\Bigg]
      \end{eqnarray}
\noindent ($h\nu_{0}/k_{B}T_{0}= \lambda_{Th0}/\lambda_{0}$).

 For wavelengths short enough that  $e^{\lambda_{Tho}/\lambda_{0}}-1\approx e^{\lambda_{Tho}/\lambda_{0}}$ 
(say for $\lambda_{0}\lesssim.25$cm, since $\lambda_{Tho}\approx .5$cm), for which the validity of (\ref{33}) is perhaps problematic (since, as mentioned, the approximation made in its evaluation is best for longer wavelengths), it is nonetheless interesting to note that 
the bracket in (\ref{34}) becomes $[1-(\lambda_{Tho}/\lambda_{0})\Delta]$.  This is identical in form to the bracket in (\ref{33}), a constant divided by $\lambda_{0}$, so  
the spectrum with the collapse diminution of photons is indistinguishable from a blackbody spectrum with a lower temperature such that 
$\Delta=.06\lambda/\lambda_{Tho}\approx .1\lambda$.  This provides no limit on $\lambda$.

For wavelengths long enough that $e^{\lambda_{Tho}/\lambda_{0}}-1\approx \lambda_{Tho}/\lambda_{0}$ (say $\lambda_{0}\gtrsim 1$cm), 
the bracket in (\ref{34}) becomes $\approx[1-\Delta]$.  Since there is not the collapse dependence $\sim1/\lambda_{0}$, for large enough 
$\lambda$ one would see an alteration of the spectrum shape that could be distinguished from a blackbody spectrum with a different temperature. Since one does not see such an alteration, if one supposes that it is there, then the error masks it.  In that case, from (\ref{34}), one has 
$.06\lambda/\lambda_{0}\lesssim\Delta$ which, for $\lambda_{0}=1$cm, gives the limitation $\lambda\lesssim 3\times 10^{-3}$s$^{-1}$.

One might do another calculation, considering that the photons lost from the blackbody radiation were converted to higher energy photons, on the wavelength scale $2\pi a$, 
and calculate the distribution of this anomalous radiation arriving along with the unaltered photons comprising the bulk of the blackbody radiation. However, there are many sources of infrared radiation\cite{infr}, among which one could not detect this contribution.

A concluding remark.

There are two reasons why the excitation effect of collapse is small on the cosmic radiation and, indeed, upon all the photon collections considered here, compared say to the excitation of electrons in atoms.  One is that the effect increases with mass-energy, and the photons considered here have so much less energy than the 
mass-energy of electrons.  The other is that one can experimentally compensate for the small rate of excitation of particles by observing many particles over a long period of time but, while it is possible to conveniently amass and observe large numbers of atoms, for a long time,  photons are not readily accumulated in large amounts, as they insist upon scurrying away.

\appendix
\section{Collapse}\label{A}

Here we apply the density matrix evolution Eq.(\ref{1}) to examine the collapse of the initial superposition state $|\psi,0\rangle =\frac{1}{\sqrt{2}} [|L\rangle +|R\rangle]$, 
of two spatially displaced states, each  consisting of $N$ superposed particles in a volume $\sim\sigma^{3}$, 
all moving with the very well-defined momentum ${\bf k}_{0}$ (we assume $k_{0}\sigma>> k_{0}a>>1$) in a direction orthogonal to the vector ${\bf x}_{L}-{\bf x}_{R}$ 
(we assume $|{\bf x}_{L}-{\bf x}_{R}|>\sigma$)
connecting them: 
\begin{equation}\label{A1}
|L\rangle\equiv  \frac{1}{\sqrt{N!}}\Big[\int d{\bf x}\xi^{\dagger}({\bf x})\frac{1}{(2\pi \sigma^{2})^{3/4}}e^{-({\bf x}-{\bf x}_{L})^{2}/4\sigma^{2}}e^{i{\bf k}_{0}\cdot{\bf x}}\Big]^{N} |0\rangle
=  \frac{1}{\sqrt{N!}}\Big[\int d{\bf k}a^{\dagger}({\bf k})\Big(\frac{2 \sigma^{2}}{\pi}\Big)^{3/4}e^{-({\bf k}-{\bf k}_{0})^{2}\sigma^{2}}e^{-{i{\bf k}\cdot{\bf x}_{L}}}\Big]^{N} |0\rangle,
\end{equation}
\noindent and similarly for $|R\rangle$, with ${\bf x}_{L}\rightarrow {\bf x}_{R}$.  We want $\langle L|R\rangle=e^{-N|{\bf x}_{L}-{\bf x}_{R}|^{2}/8\sigma^{2}}\approx 0$,  so we 
assume  $N|{\bf x}_{L}-{\bf x}_{R}|^{2}>>8\sigma^{2}$.
We shall denote 
$\alpha_{L}^{*}({\bf k})=\Big(\frac{2 \sigma^{2}}{\pi}\Big)^{3/4}e^{-({\bf k}-{\bf k}_{0})^{2}\sigma^{2}}e^{-{i{\bf k}\cdot{\bf x}_{L}}}$ and $\omega(k)\equiv\sqrt{k^{2}+M^{2}}$.

We note that each of these states is an approximate energy eigenstate to high accuracy:
\begin{equation}\label{A2}
H|L\rangle=\int d{\bf k}\omega(k)a^{\dagger}({\bf k})a({\bf k})|L\rangle=
N\frac{1}{\sqrt{N!}}\Big[\int d{\bf k}a^{\dagger}({\bf k})\alpha_{L}^{*}({\bf k})\Big]^{N-1} 
\int d{\bf k}a^{\dagger}({\bf k})\omega( k)\alpha_{L}^{*}({\bf k})|0\rangle
\approx N\omega( k_{0})|L\rangle,
\end{equation}
\noindent since the gaussian in $\alpha({\bf k}) $ implies $k=k_{0}+o(1/\sigma)$.   

We wish to show that the off-diagonal elements of the density matrix decay, signaling  collapse to one or the other state,  to first order in $\lambda$. When we 
calculate $\langle L|\rho(t)|R \rangle$ using (\ref{1}), the Hamiltonian term makes no contribution since 
$\langle L|[H,\rho(t)]|R \rangle\approx N\omega({\bf k}_{0})[\langle L|\rho(t)|R\rangle-\langle L|\rho(t)|R\rangle]=0$, so we get, with 
$\rho(0)=\frac{1}{2}[|L\rangle +|R\rangle][\langle L| +\langle R|]$ (and noting that the matrix elements of the energy density operators between $\langle L|$ and $|R\rangle$ essentially vanish because the overlap integral between left and right states when there is one or two less particles is still negligibly small): 
\begin{eqnarray}\label{A3}
\langle L|\rho(t)|R\rangle&=& \frac{1}{2}-\frac{\lambda t}{2M_{N}^{2}}\int d{\bf x}\int d{\bf x}'e^{-({\bf x}-{\bf x}')^{2}/4a^{2}}\nonumber\\
&&\Big[\langle L| K^{1/2}\xi^{\dagger}({\bf x})K^{1/2}\xi({\bf x})K^{1/2}\xi^{\dagger}({\bf x}')K^{1/2}\xi({\bf x}')|L\rangle\frac{1}{2}+
\frac{1}{2}\langle R| K^{1/2}\xi^{\dagger}({\bf x})K^{1/2}\xi({\bf x})K^{1/2}\xi^{\dagger}({\bf x}')K^{1/2}\xi({\bf x}')|R\rangle\nonumber\\
&&\qquad\qquad\qquad\qquad-2\frac{1}{2}\langle L|K^{1/2}\xi^{\dagger}({\bf x})K^{1/2}\xi({\bf x})|L\rangle\langle R|K^{1/2}\xi^{\dagger}({\bf x}')K^{1/2}\xi({\bf x}')|R\rangle\Big]\nonumber\\
&=& \frac{1}{2}-\frac{\lambda t}{4M_{N}^{2}}\Big(\frac{a^{2}}{\pi}\Big)^{3/2}\int d{\bf k}_{1}d{\bf k}_{2}d{\bf k}_{3}d{\bf k}_{4}\sqrt{ \omega(k_{1})\omega(k_{2})\omega( k_{3} )\omega(k_{4})}e^{-({\bf k}_{1}-{\bf k}_{2})^{2}a^{2}}\delta(-{\bf k}_{1}+{\bf k}_{2}-{\bf k}_{3}+{\bf k}_{4})\nonumber\\
&&\Big[\langle L| a^{\dagger}({\bf k}_{1})[a^{\dagger}({\bf k}_{3})a({\bf k}_{2})+\delta({\bf k}_{2}-{\bf k}_{3})]a({\bf k}_{4})|L\rangle+
\langle R| a^{\dagger}({\bf k}_{1})[a^{\dagger}({\bf k}_{3})a({\bf k}_{2})+\delta({\bf k}_{2}-{\bf k}_{3})]a({\bf k}_{4})|R\rangle\nonumber\\
&&\qquad\qquad\qquad\qquad-2\langle L|a^{\dagger}({\bf k}_{1})a({\bf k}_{2})|L\rangle\langle R|a^{\dagger}({\bf k}_{3})a({\bf k}_{4})|R\rangle\Big]\nonumber\\
&=&\frac{1}{2}-\frac{\lambda t}{4M_{N}^{2}}\Big(\frac{a^{2}}{\pi}\Big)^{3/2}\int d{\bf k}_{1}d{\bf k}_{2}d{\bf k}_{3}d{\bf k}_{4}\sqrt{ \omega(k_{1})\omega(k_{2})\omega( k_{3} )\omega(k_{4})}e^{-({\bf k}_{1}-{\bf k}_{2})^{2}a^{2}}\delta(-{\bf k}_{1}+{\bf k}_{2}-{\bf k}_{3}+{\bf k}_{4})\nonumber\\
&&2\Big[N(N-1)\alpha({\bf k}_{1})\alpha({\bf k}_{2})\alpha({\bf k}_{3})\alpha({\bf k}_{4})+ N\alpha({\bf k}_{1})\alpha({\bf k}_{4})\delta({\bf k}_{2}-{\bf k}_{3}) 
-N^{2}a_{L}^{*}({\bf k}_{1})\alpha_{L}({\bf k}_{2})\alpha_{R}^{*}({\bf k}_{3})\alpha_{R}({\bf k}_{4})\Big],
\end{eqnarray}
\noindent where we have set $\alpha({\bf k)}\equiv|\alpha_{L,R}({\bf k)}|$.

We shall now evaluate the three integrals involving the bracketed terms in Eq.(\ref{A3}). 

For the first term, again, we utilize the excellent approximation $f({\bf k})\alpha({\bf k})\approx f({\bf k}_{0})\alpha({\bf k})$ (since 
the factor  $\alpha({\bf k})$ implies ${\bf k}={\bf k}_{0}$ to order $1/\sigma$):
\begin{eqnarray}\label{A4}
I_{1}&\equiv&\Big(\frac{a^{2}}{\pi}\Big)^{3/2}\int d{\bf k}_{1}d{\bf k}_{2}d{\bf k}_{3}d{\bf k}_{4}\sqrt{ \omega(k_{1})\omega(k_{2})\omega( k_{3} )\omega(k_{4})}e^{-({\bf k}_{1}-{\bf k}_{2})^{2}a^{2}}\delta(-{\bf k}_{1}+{\bf k}_{2}-{\bf k}_{3}+{\bf k}_{4})
\alpha({\bf k}_{1})\alpha({\bf k}_{2})\alpha({\bf k}_{3})\alpha({\bf k}_{4})\nonumber\\
&\approx&\omega^{2}(k_{0})\Big(\frac{a^{2}}{\pi}\Big)^{3/2}\int d{\bf k}_{1}d{\bf k}_{2}d{\bf k}_{3}d{\bf k}_{4}\delta(-{\bf k}_{1}+{\bf k}_{2}-{\bf k}_{3}+{\bf k}_{4})
\alpha({\bf k}_{1})\alpha({\bf k}_{2})\alpha({\bf k}_{3})\alpha({\bf k}_{4})\nonumber\\
&=&\omega^{2}(k_{0})\Big(\frac{a^{2}}{\pi}\Big)^{3/2}\int d{\bf k}_{1}d{\bf k}_{2}d{\bf k}_{3}d{\bf k}_{4}\frac{1}{(2\pi)^{3}}
\int d{\bf x}e^{i{\bf x}\cdot (-{\bf k}_{1}+{\bf k}_{2}-{\bf k}_{3}+{\bf k}_{4})}
\alpha({\bf k}_{1})\alpha({\bf k}_{2})\alpha({\bf k}_{3})\alpha({\bf k}_{4})\nonumber\\
&=&\omega^{2}(k_{0})\Big(\frac{a^{2}}{\pi}\Big)^{3/2}\frac{1}{(2\pi)^{3}}
\int d{\bf x}\Big(\frac{2\sigma^{2}}{\pi}\Big)^{3}\Big[\Big(\frac{\pi}{\sigma^{2}}\Big)^{3/2}e^{-{\bf x}^{2}/4\sigma^{2}}\Big]^{4}\nonumber\\
&=&\omega^{2}(k_{0})\Big(\frac{a}{\sigma}\Big)^{3}.
\end{eqnarray}

The second integral is, using the same approximation,
\begin{eqnarray}\label{A5}
I_{2}&\equiv&\Big(\frac{a^{2}}{\pi}\Big)^{3/2}\int d{\bf k}_{1}d{\bf k}_{2}d{\bf k}_{3}d{\bf k}_{4}\sqrt{ \omega(k_{1})\omega(k_{2})\omega( k_{3} )\omega(k_{4})}e^{-({\bf k}_{1}-{\bf k}_{2})^{2}a^{2}}\delta(-{\bf k}_{1}+{\bf k}_{2}-{\bf k}_{3}+{\bf k}_{4})\alpha({\bf k}_{1})\alpha({\bf k}_{4})\delta({\bf k}_{2}-{\bf k}_{3}) \nonumber\\
&=&\Big(\frac{a^{2}}{\pi}\Big)^{3/2}\int d{\bf k}_{1}d{\bf k}_{2}\omega(k_{1})\omega(k_{2})e^{-({\bf k}_{1}-{\bf k}_{2})^{2}a^{2}}\alpha^{2}({\bf k}_{1})\nonumber\\
&\approx&\omega(k_{0})\Big(\frac{a^{2}}{\pi}\Big)^{3/2}\int d{\bf k}_{1}d{\bf k}_{2}\omega(k_{2})e^{-({\bf k}_{0}-{\bf k}_{2})^{2}a^{2}}\alpha^{2}({\bf k}_{1}).
\end{eqnarray}
\noindent We may likewise use the approximation $\omega(k_{2})e^{-({\bf k}_{0}-{\bf k}_{2})^{2}a^{2}}\approx \omega(k_{0})e^{-({\bf k}_{0}-{\bf k}_{2})^{2}a^{2}}$, since 
the gaussian implies ${\bf k}_{2}={\bf k}_{0}$ to order $1/a$. Then, 
\begin{eqnarray}\label{A6}
I_{2}&\approx&
\omega^{2}(k_{0})\Big(\frac{a^{2}}{\pi}\Big)^{3/2}\int d{\bf k}_{1}d{\bf k}_{2}e^{-({\bf k}_{0}-{\bf k}_{2})^{2}a^{2}}
\Big(\frac{2 \sigma^{2}}{\pi}\Big)^{3/2}e^{-({\bf k}_{1}-{\bf k}_{0})^{2}2\sigma^{2}}\nonumber\\
&=&\omega^{2}(k_{0}).
\end{eqnarray}

The last integral, with the already employed approximations, is
\begin{eqnarray}\label{A7}
I_{3}&\approx&\omega^{2}(k_{0})\Big(\frac{a^{2}}{\pi}\Big)^{3/2}\int d{\bf k}_{1}d{\bf k}_{2}d{\bf k}_{3}d{\bf k}_{4}\delta(-{\bf k}_{1}+{\bf k}_{2}-{\bf k}_{3}+{\bf k}_{4})\alpha({\bf k}_{1})\alpha({\bf k}_{2})\alpha({\bf k}_{3})\alpha({\bf k}_{4})
e^{i{\bf x}_{L}\cdot({\bf k}_{2}-{\bf k}_{1})}e^{i{\bf x}_{R}\cdot({\bf k}_{4}-{\bf k}_{3})}\nonumber\\
&=&\omega^{2}(k_{0})\Big(\frac{a^{2}}{\pi}\Big)^{3/2}\frac{1}{(2\pi)^{3}}
\int d{\bf x}\int d{\bf k}_{1}d{\bf k}_{2}d{\bf k}_{3}d{\bf k}_{4}e^{i{\bf x}\cdot (-{\bf k}_{1}+{\bf k}_{2}-{\bf k}_{3}+{\bf k}_{4})}\alpha({\bf k}_{1})\alpha({\bf k}_{2})\alpha({\bf k}_{3})\alpha({\bf k}_{4})e^{i({\bf x}_{L}-{\bf x}_{R})\cdot({\bf k}_{2}-{\bf k}_{1})}\nonumber\\
&=&\omega^{2}(k_{0})\Big(\frac{a^{2}}{\pi}\Big)^{3/2}\frac{1}{(2\pi)^{3}}
\int d{\bf x}\int d{\bf k}_{1}d{\bf k}_{2}e^{i{\bf x}\cdot (-{\bf k}_{1}+{\bf k}_{2})}\alpha({\bf k}_{1})\alpha({\bf k}_{2})e^{i({\bf x}_{L}-{\bf x}_{R})\cdot({\bf k}_{2}-{\bf k}_{1})}
\Big(\frac{2\pi}{\sigma^{2}}\Big)^{3/2}e^{-\frac{{\bf x}^{2}}{2\sigma^{2}}}\nonumber\\
&=&\omega^{2}(k_{0})\Big(\frac{a^{2}}{\pi}\Big)^{3/2}
\int d{\bf k}_{1}d{\bf k}_{2}\alpha({\bf k}_{1})\alpha({\bf k}_{2})e^{i({\bf x}_{L}-{\bf x}_{R})\cdot({\bf k}_{2}-{\bf k}_{1})}e^{-({\bf k}_{1}-{\bf k}_{2})^{2}\sigma^{2}/2}.
\end{eqnarray}
\noindent Upon making the replacement ${\bf k}'_{i}=({\bf k}_{i}-{\bf k}_{0})$, and then removing the primes, we continue:
\begin{eqnarray}\label{A8}
I_{3}
&=&\omega^{2}(k_{0})\Big(\frac{a^{2}}{\pi}\Big)^{3/2}\Big(\frac{2\sigma^{2}}{\pi}\Big)^{3/2}
\int d{\bf k}_{1}d{\bf k}_{2}e^{-{\bf k}_{1}^{2}\sigma^{2}}e^{-{\bf k}_{2}^{2}\sigma^{2}}e^{i({\bf x}_{L}-{\bf x}_{R})\cdot({\bf k}_{2}-{\bf k}_{1})}e^{-({\bf k}_{1}-{\bf k}_{2})^{2}\sigma^{2}/2}
\nonumber\\
&=&\omega^{2}(k_{0})\Big(\frac{a^{2}}{\pi}\Big)^{3/2}\Big(\frac{2\sigma^{2}}{\pi}\Big)^{3/2}
\int d{\bf k}_{1}d{\bf k}_{2}e^{-({\bf k}_{1}+{\bf k}_{2})^{2}\sigma^{2}/2}e^{i({\bf x}_{L}-{\bf x}_{R})\cdot({\bf k}_{2}-{\bf k}_{1})}e^{-({\bf k}_{1}-{\bf k}_{2})^{2}\sigma^{2}}.
\end{eqnarray}
With change of variables ${\bf k}'\equiv{\bf k}_{1}+{\bf k}_{2}, {\bf k}\equiv{\bf k}_{1}-{\bf k}_{2}$ and so $d{\bf k}_{1}{d\bf k}_{2}=\frac{1}{8}d{\bf k}d{\bf k}'$:
\begin{eqnarray}\label{A9}
I_{3}
&=&\omega^{2}(k_{0})\Big(\frac{a^{2}}{\pi}\Big)^{3/2}\Big(\frac{\sigma^{2}}{2\pi}\Big)^{3/2}
\int d{\bf k}d{\bf k}'e^{-{\bf k}'^{2}\sigma^{2}/2}e^{-i({\bf x}_{L}-{\bf x}_{R})\cdot{\bf k}}e^{-{\bf k}^{2}\sigma^{2}}\nonumber\\
&=&\omega^{2}(k_{0})\Big(\frac{a}{\sigma}\Big)^{3}e^{-({\bf x}_{L}-{\bf x}_{R})^{2}/4\sigma^{2}}.
\end{eqnarray}

Putting (\ref{A4}), (\ref{A6}) and (\ref{A9}) into (\ref{A3}), we obtain the expression for the off-diagonal matrix element, 
\begin{eqnarray}\label{A10}
\langle L|\rho(t)|R\rangle 
&\approx&\frac{1}{2}-\frac{N\lambda t\omega^{2}(k_{0})}{2M_{N}^{2}}\Big[N\Big(\frac{a}{\sigma}\Big)^{3}\Big(1-e^{-({\bf x}_{L}-{\bf x}_{R})^{2}/4\sigma^{2}}\Big)
+\Big(1-\Big(\frac{a}{\sigma}\Big)^{3}\Big)\Big]\nonumber\\
&\approx&\frac{1}{2}-\frac{N\lambda t\omega^{2}(k_{0})}{2M_{N}^{2}}\Big[N\Big(\frac{a}{\sigma}\Big)^{3}\Big(1-e^{-({\bf x}_{L}-{\bf x}_{R})^{2}/4\sigma^{2}}\Big)
+1\Big],
\end{eqnarray}
\noindent cited in Section I.

\section {Integrals} \label{B}
\subsection{Integral involved in the first bracketed term in Eq.(\ref{14}).}

We wish to evaluate the integral 
\begin{eqnarray}\label{B1}
I&\equiv&\int d{\bf k}_{1}d{\bf k}_{2}d{\bf k}_{3}d{\bf k}_{4}\sqrt{ k_{1}k_{2} k_{3} k_{4}}e^{-({\bf k}_{1}-{\bf k}_{2})^{2}a^{2}}\delta(-{\bf k}_{1}+{\bf k}_{2}-{\bf k}_{3}+{\bf k}_{4})
\alpha({\bf k}_{1})\alpha({\bf k}_{2})\alpha({\bf k}_{3})\alpha({\bf k}_{4})\nonumber\\
&\approx&k_{0}^{2}\int d{\bf k}_{1}d{\bf k}_{2}d{\bf k}_{3}d{\bf k}_{4}e^{-({\bf k}_{1}-{\bf k}_{2})^{2}a^{2}}\frac{1}{(2\pi)^{3}}\int d{\bf x}
e^{i{\bf x}\cdot[-{\bf k}_{1}+{\bf k}_{2}-{\bf k}_{3}+{\bf k}_{4})]}\nonumber\\
&&\frac{1}{(\pi/\sigma^{2})^{3}}e^{-({\bf k}_{1}-{\bf k}_{0})^{2}\sigma^{2}/2}e^{-({\bf k}_{2}-{\bf k}_{0})^{2}\sigma^{2}/2}e^{-({\bf k}_{3}-{\bf k}_{0})^{2}\sigma^{2}/2}e^{-({\bf k}_{4}-{\bf k}_{0})^{2}\sigma^{2}/2}
\end{eqnarray}
\noindent We make the change of variables ${\bf k}'_{i}\equiv{\bf k}_{i}-{\bf k}_{0}$, thereafter unpriming the ${\bf k}$'s, and perform the integrals over ${\bf k}_{3}, {\bf k}_{4}$:
\begin{eqnarray}\label{B2}
I&\equiv&k_{0}^{2}\int d{\bf k}_{1}d{\bf k}_{2}
e^{-({\bf k}_{1}-{\bf k}_{2})^{2}a^{2}}\frac{1}{(2\pi)^{3}}\int d{\bf x}
e^{i{\bf x}\cdot[-{\bf k}_{1}+{\bf k}_{2}]}\Bigg(\frac{2\pi}{\sigma^{2}}\Bigg)^{3}e^{-{\bf x}^{2}/\sigma^{2}}
\frac{1}{(\pi/\sigma^{2})^{3}}e^{-{\bf k}_{1}^{2}\sigma^{2}/2}e^{-{\bf k}_{2}^{2}\sigma^{2}/2}.\nonumber\\
\end{eqnarray}
\noindent Changing variables to ${\bf k}\equiv{\bf k}_{1}-{\bf k}_{2}, {\bf k}'\equiv{\bf k}_{1}+{\bf k}_{2}$, using $d{\bf k}d{\bf k}'=8d{\bf k}_{1}d{\bf k}_{2}$:
\begin{eqnarray}\label{B3}
I&\equiv&
\frac{1}{(2\pi)^{3}}k_{0}^{2}\int d{\bf x}
e^{-{\bf x}^{2}/\sigma^{2}}\int d{\bf k}e^{-{\bf k}^{2}(a^{2}+\sigma^{2}/4)}e^{-i{\bf x}\cdot{\bf k}}\int d{\bf k}'e^{-{\bf k}'^{2}\sigma^{2}/4}.
\end{eqnarray}
\noindent Since $\sigma>>a$, we can make the approximation  $a^{2}+\sigma^{2}/4\approx \sigma^{2}/4$, and so obtain:
\begin{eqnarray}\label{B4}
I&\approx&
\frac{1}{(2\pi)^{3}}k_{0}^{2}\int d{\bf x}
e^{-{\bf x}^{2}/\sigma^{2}}\Bigg(\frac{4\pi}{\sigma^{2}}\Bigg)^{3/2}e^{-{\bf x}^{2}/\sigma^{2}}\Bigg(\frac{4\pi}{\sigma^{2}}\Bigg)^{3/2}=k_{0}^{2}\frac{(2\pi)^{3/2}}{\sigma^{3}}.
\end{eqnarray}
\noindent As noted, (\ref{B4}) $\sim\frac{k_{0}^{2}}{\sigma^{3}}$  is small compared to the second bracketed term in Eq.(\ref{14}) which is, according to (\ref{15a}) or (\ref{15b}) 
$\sim\frac{k_{0}}{a^{4}}$ or $\sim\frac{k_{0}^{2}}{a^{3}}$,
and so (\ref{B4}) may be neglected. 

\subsection{Integrals involved in Eq.(\ref{21}).}

We wish to evaluate the integral
\begin{eqnarray}\label{B5}
I'&\equiv&\int d{\bf k}_{1}d{\bf k}_{2}d{\bf k}_{3}d{\bf k}_{4}\sqrt{ k_{1}k_{2} k_{3} k_{4}}e^{-({\bf k}_{1}-{\bf k}_{2})^{2}a^{2}}\delta(-{\bf k}_{1}+{\bf k}_{2}-{\bf k}_{3}+{\bf k}_{4})
\alpha({\bf k}_{1})\alpha({\bf k}_{2})\alpha({\bf k}_{3})\alpha({\bf k}_{4})\nonumber\\
&&\qquad\qquad\qquad\qquad\cdot\Bigg[\alpha^{2}({\bf k})-2\delta({\bf k}_{2}-{\bf k}))\Bigg]\nonumber\\
\end{eqnarray}\label{B6}
The integral multiplying $\alpha^{2}({\bf k})$ is the result (\ref{B4}), so the first bracketed term in (\ref{B4}) leads to 
\begin{equation}
I_{1}=k_{0}^{2}2^{3/2}e^{-({\bf k}-{\bf k}_{0})^{2}\sigma^{2}}.
\end{equation}
As for the second bracketed term in (\ref{B5}), this is the integral (\ref{B2}) with $-2\delta({\bf k}_{2}-{\bf k})$ inserted: 
\begin{eqnarray}\label{B7}
I_{2}&\equiv&-2\frac{1}{\pi^{3}}k_{0}^{2}\int d{\bf k}_{1}
e^{-({\bf k}_{1}-{\bf k})^{2}a^{2}}\int d{\bf x}
e^{i{\bf x}\cdot[-{\bf k}_{1}+{\bf k}]}e^{-{\bf x}^{2}/\sigma^{2}}
e^{-{\bf k}_{1}^{2}\sigma^{2}/2}e^{-{\bf k}^{2}\sigma^{2}/2}\nonumber\\
&=&-2\Big(\frac{\sigma^{2}}{\pi}\Big)^{3/2}k_{0}^{2}\int d{\bf k}_{1}
e^{-({\bf k}_{1}-{\bf k})^{2}(a^{2}+\sigma^{2}/4)}
e^{-{\bf k}_{1}^{2}\sigma^{2}/2}e^{-{\bf k}^{2}\sigma^{2}/2}\nonumber\\
&\approx&-2\Big(\frac{\sigma^{2}}{\pi}\Big)^{3/2}k_{0}^{2}\int d{\bf k}_{1}
e^{-({\bf k}_{1}-{\bf k})^{2}\sigma^{2}/4}
e^{-{\bf k}_{1}^{2}\sigma^{2}/2}e^{-{\bf k}^{2}\sigma^{2}/2}\nonumber\\
&=&-2k_{0}^{2}\Big(\frac{4}{3}\Big)^{3/2} 
e^{-({\bf k}-{\bf k}_{0})^{2}2\sigma^{2}/3}.
\end{eqnarray}

 \end{document}